# Enhance CNN Robustness Against Noises for Classification of 12-Lead ECG with Variable Length

Linhai Ma
Department of Computer Science
University of Miami
Coral Gables, USA
l.ma@miami.edu

Liang Liang
Department of Computer Science
University of Miami
Coral Gables, USA
liang.liang@miami.edu

*Abstract*— Electrocardiogram (ECG) is the most widely used diagnostic tool to monitor the condition of the cardiovascular system. Deep neural networks (DNNs), have been developed in many research labs for automatic interpretation of ECG signals to identify potential abnormalities in patient hearts. Studies have shown that given a sufficiently large amount of data, the classification accuracy of DNNs could reach human-expert cardiologist level. However, despite of the excellent performance in classification accuracy, it has been shown that DNNs are highly vulnerable to adversarial noises which are subtle changes in input of a DNN and lead to a wrong class-label prediction with a high confidence. Thus, it is challenging and essential to improve robustness of DNNs against adversarial noises for ECG signal classification –a life-critical application. In this work, we designed a CNN for classification of 12-lead ECG signals with variable length, and we applied three defense methods to improve robustness of this CNN for this classification task. The ECG data in this study is very challenging because the sample size is limited, and the length of each ECG recording varies in a large range. The evaluation results show that our customized CNN reached satisfying F1 score and average accuracy, comparable to the top-6 entries in the CPSC2018 ECG classification challenge, and the defense methods enhanced robustness of our CNN against adversarial noises and white noises, with a minimal reduction in accuracy on clean data.

## I. INTRODUCTION

Electrocardiogram (ECG) is widely used for monitoring the condition of cardiovascular system. After many years of residency training, a cardiologist becomes experienced in reading ECG graphs, detect abnormalities, and classify signals into different disease categories, which is tedious and time-consuming. As an alternative solution, researchers have developed deep neural networks (DNNs), especially convolutional neural networks (CNNs) for ECG signal analysis with excellent classification accuracy [1]. S. Kiranyaz et al. [2] designed an adaptive one-dimensional (1D) CNN for ECG classification, and the experiment showed that their method can achieve not only a high classification accuracy (99%) but also good performance in sensitivity (95.9%), specificity (99.5%), and positive predictivity (96.2 %). In 2017, U. Rajendra et al. [3] applied a CNN to classify the heartbeats, and their model achieved an accuracy of 94.03% and 93.47% for the diagnostic classification of heartbeats in ECGs, respectively. M. Kachuee et al. [4] designed a residual CNN to classify ECG heartbeats into five categories, and the CNN model was trained and evaluated on PhysioNet's MIT-BIH datasets and achieved an accuracy of 93.4%. Awni Y. Hannun, et al. [5] developed a CNN to classify 12 rhythm classes using a dataset of 91,232 single-lead ECGs from 53,877 patients, which is the largest (yet private) dataset compared to other datasets in the literature, and the classification accuracy was similar to that of cardiologists. Therefore, with a sufficiently large amount of data and a carefully designed network structure, DNN models could reach human expert radiologist level for ECG signal classification, and for each patient, the analysis can be done in a fraction of a second. For patients in developing countries where human-expert radiologists are lacking, a DNN-based automated ECG diagnostic system would be an affordable solution to improve health outcomes.

However, recent studies have shown that despite the high classification accuracy of DNNs, they are susceptible to adversarial noises, which are small perturbations to input of the networks, even imperceptible to human eyes, and able to change the prediction of DNNs [6]. Adversarial noises are usually generated by algorithms, called adversarial attacks, which can be classified into two categories based on whether the whole structure of the network is known by the attacker. It is a white box adversarial attack if the attacker knows the inner structure of the network. White box attacks often use the gradient information (e.g. gradient of the loss with respect to input) from the target network to construct adversarial noise which is added to the original input, and two well-known white box attacks are Fast Gradient Signed Method (FGSM) [7] and Projected Gradient Descent (PGD) [8]. It is a black box attack if the attacker has almost no knowledge of the inner structure of the network. One approach for black box attack is called transfer-based attack, in which a substitute network is designed and trained on the input-output pairs of the oracle network (the target network to be attacked), so that the substitute network can learn the decision boundary of the oracle network; after training, the substitute network can be used to construct adversarial noises to attack the oracle network [9]. Even simpler, the substitute network can be trained on the dataset that is used to train the oracle, and then adversarial noises are constructed using the substitute network and subsequently applied to the oracle network, with the hope that the adversarial noises are transferrable between networks. The weakness of transfer-based attacks is discussed by Jonathan Uesato et. al in [10]: "success of the attack is highly dependent on the similarity between" the substitute and the oracle. Another approach is to generate random samples around the input sample and use these random samples together with the original sample to estimate the gradient of the target network, and then the estimated gradient can be used for attacks [10]. These attacks pose significant threats [6] to the deep learning systems in sensitive and life-critical application fields such as ECG classification.

To improve DNN robustness against adversarial noises, lots of effort has been made by researchers to develop defense methods. Currently, the most popular defense strategy is adversarial training. The basic idea of adversarial training is to generate adversarial noises by using an adversarial attack and add the noises to the samples for training. The resulting noisy samples are called adversarial samples. Through adversarial training, the network can learn some features of adversarial noises, and then its decision boundary is updated so that it will become difficult to push the input across the decision boundary by adding a small amount of noise. Adversarial training is straightforward but has problems. For example, it is very computationally expensive and time-consuming to generate adversarial samples, and low-quality adversarial samples can be misleading and even reduce the classification accuracy of networks. Therefore, different adversarial training-based defense methods were proposed [6], which share the same basic idea and vary in how the adversarial samples are generated. Ali et al. [11] designed an efficient algorithm to reduce the high computation cost of generating adversarial samples, which needs to alter the standard training process. Zhang et al. [12] proposed to generate high-quality adversarial examples with less affection on the accuracy of network. Parallel to adversarial training, regularization terms can be added to the loss function to reduce the sensitivity of network output with respect to the input. Regularization terms could be the gradient magnitude of loss with respect to input [13], Jacobian regularization [14], or NSR Regularization [15].

In this paper, we designed a CNN for classification of ECG signals and applied three defense methods to improve robustness of this CNN against adversarial noises as well as white noises. One of the defense methods is based on adversarial training, and the other two methods uses regularization terms. The ECG data in experiment is publicly available from the China Physiological Signal Challenge 2018 (CPSC2018) [16]. The dataset is challenging for classification. The number of recordings is only 6,877, and the time duration of a recording can vary from 6 to 60 seconds. With a sampling rate of 500 Hz, the number of amplitude values in a digital ECG signal can vary from 3000 to 30000. Results from our experiment showed that our customized CNN reached satisfying F1 score and accuracy, compared to top-6 methods in CPSC2018, and the defense methods successfully enhanced robustness of this CNN against adversarial noises and white noises, with a minimal reduction in accuracy on clean data.

## II. METHODOLOY

In this section, we will present our customized CNN, discuss the adversarial robustness issue, and introduce the three defense methods that we applied to improve the robustness of the CNN in this study.

### A. The Customized CNN

We designed a CNN to classify the ECG signals and handle the challenge of variable input signal length. The high-level architecture of this neural network is shown in Fig. 1. To describe the CNN, we use "Cov (a, b, c, d, e)" to represent a convolution layer with "a" input channels, "b" output channels, kernel size of "c", stride of "d" and padding of "e"; "Linear (f, g)" to denote a linear layer with input size of "f" and output size of "g"; "MaxPool (h, i, j)" to denote max-pooling with kernel size of h, stride of i and padding size of j; "AvgPool (k, l, m)" to denote average-pooling with kernel size of k, stride of l and padding size of m; "GN" to denote group normalization [17]. The network has 4 convolution blocks of the same structure as shown in Fig. 1, which will shrinkage the size of the input tensor and doubled the number of channels.

RNN is typical used to handle a time sequence with variable length. However, RNN runs very slow. To enable the CNN to handle variable signal length, we used zero-padding and mask. Each signal (i.e. the ECG recording from a patient) is padded with zeros to a fixed length of 33792 in order to process the signals in mini-batches. The network should ignore the zeros in signals, and this is realized by using an input mask with a fixed length of 33792. For valid elements of the padded signal, the corresponding elements of the mask are ones. For zero elements of the padded signal, the corresponding elements of the mask are zeros. To obtain the output mask, an average-pooling is applied to the input mask by estimating what the mask should be after a series of convolutional operations on the mask. Then, a valid output feature is obtained by multiplying the output mask and the feature output from the convolution blocks. After this operation, a channel-wise averaging weighted by the mask is performed to reduce the dimension of the feature from a variable length to a fixed length of 512. After the final linear layer, classification scores (a.k.a. logits) are obtained.

### B. PGD-based Adversarial Attack and Adversariel Training

Projected Gradient Descent (PGD) [8] is regarded as the strongest first-order white box adversarial attack. To evaluate the robustness of different adversarial training methods [8][18], we use projected gradient descent (PGD) to generate adversarial noises, which is widely used for defense method evaluation [10][19]. K-PGD attack on clean input signal $x$ with $K$ iterations are performed by:

$$x^k = \prod \left( x^{k-1} + \alpha \cdot sign(\nabla_X Loss(x^{k-1})) \right) \quad (1)$$

where $\alpha$ is the step size and $x^k$ is the adversarial example from the $k$-th iteration/step. If the noise added to input $x$ is larger than the given noise level $\epsilon$, the projection operation $\prod()$ in PGD will project it back onto a $\epsilon$-ball, i.e, $\|x^k - x\|_p \leq \epsilon$ where $\| \ \|_p$ is the vector Lp norm and p is usually infinity or 2. Assuming $x$ is correctly-classified by a network, then, after

Identify applicable funding agency here. If none, delete this text box.

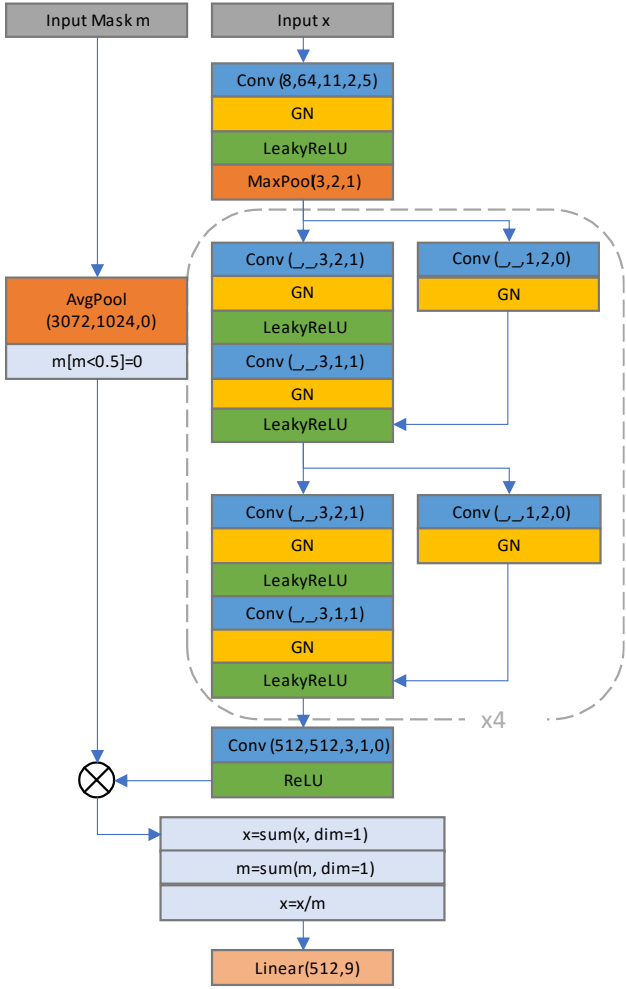

Fig. 1. The architecture of the CNN

$K$ iterations, the $Loss$ on the noisy sample $x^K$ becomes very large such that $x^K$ is wrong-classified. Usually the decision boundary of a network, which is trained with cross-entropy loss and clean data, is very close to the samples. As a result, a tiny amount of adversarial noises on the input sample can cause a wrong classification.

In this paper, $\epsilon$ is named the noise level. We only focus on the noise level no larger than 0.01 because when $\epsilon$ is larger than 0.01, the perturbation on ECG signal is so significant that even human eyes can spot the noises on $x^K$, and it is also unrealistic to have such large noises in signals. (see Appendix for examples).

The basic idea of adversarial training [7] is to generate adversarial samples by an algorithm (e.g. PGD) and use them as part of the training data. In this way, the trained model may become robust against adversarial noises. Madry et al. [8] proposed to train models with adversarial samples produced by PGD adversarial attack and achieved state-of-the-art robustness against strong L-infinity norm based attacks on MNIST and CIFAR-10 dataset. Since the success of this PGD-based adversarial training method, many other defense methods leveraging PGD adversarial attack to generate adversarial samples have been proposed, such as Max-Margin Adversarial (MMA) Training [20] and Increasing-Margin Adversarial (IMA) Training [21], which may work better for large adversarial noises. In this experiment, we applied the vanilla adversarial training with 20-PGD [11][7] to improve the robustness of the CNN for the classification of ECG signals. The loss function of this vanilla adversarial training method is:

$$loss = 0.5 L_{CE}(x, y) + 0.5 L_{CE}(x_\epsilon, y) \quad (2)$$

where $L_{CE}$ is cross-entropy loss and $x_\epsilon$ is an adversarial sample (i.e. output from 20-PGD) on noise level of $\|\epsilon\|$. To strengthen the vanilla adversarial training methods, the level of the noise added to the input is scaling up linearly as the number of epochs increases during training [20]:

$$\epsilon_t = \epsilon(t - 10)/(t_{max} - 10) \quad (3)$$

where $\epsilon_t$ is the noise level for the current epoch $t$; $\epsilon$ is the user-defined maximum noise level; $t_{max}$ is the user-defined total number of training epochs. To help the model converge, during the training process, the term $0.5 L_{CE}(x_\epsilon, y)$ will not be added to the loss function until the 11-th epoch.

*C. Jacobian Regularization*

The idea of Jacobian Regularization [14] is to penalize large gradient of the loss function with respect to the input, during the post-processing training phase. Adversarial noises are essentially small changes in the input, which cause large changes in the network's output. To reduce the sensitivity to input change, a regularization term in the form of the Frobenius norm of the network's Jacobian matrix evaluated on the input can be added to the loss function. The loss function is given by:

$$loss = L_{CE}(x, y) + \lambda \sqrt{\sum_{d=1}^{D}\sum_{k=1}^{K}\sum_{n=1}^{N}\left(\frac{\partial}{\partial x_d} z_k(x_n)\right)^2} \quad (4)$$

where $L_{CE}$ is cross-entropy loss, $D$ is the number of input dimensions, $K$ is the number of classes, $N$ is the batch size, $x = \{x_1, x_2, \ldots, x_i, \ldots, x_N\}$ is the input batch data ($n$ is sample index in the batch), $y$ contains the ground truth labels, and $z = \{z_1, z_2, \ldots, z_i, \ldots, z_K\}$ is the output logits of the neural network. $x_d$ refers to the d-th dimension of a sample $x_n$. In this experiment, because of high input dimension, we use mean instead of sum in the regularization term. In this way, we can avoid too large regularization during the training process, which makes it easier to reach the balance between robustness and accuracy. The loss function used in our experiment is

$$loss = L_{CE}(x, y) + \frac{\lambda}{NK}\sqrt{\sum_{d=1}^{D}\sum_{k=1}^{K}\sum_{n=1}^{N}\left(\frac{\partial}{\partial x_d} z_k(x_i)\right)^2} \quad (5)$$

The only parameter needs to be adjusted is $\lambda$. To help the model converge during the training process, the regularization term will not be added to loss function until the 11-th epoch.

*D. NSR Regularization*

The NSR Regularization in this study is the "Loss2" proposed in [15]. The idea of this regularization is to minimize the noise-to-signal ratio (NSR). Given an input sample $x$ (converting $x$ to a vector), the output of the CNN can be exactly expressed by a "linear" equation [20]:

$$z = W^T x + b \quad (6)$$

where the weight matrix $W$ will be different for different $x$, and the bias vector $b$ will different for different $x$. The output is a vector $z = [z_1, ..., z_9]^T$ where $z_i$ is the output logit of class-$i$ and the total number of classes is 9 in this application. Let $w_i$ be the $i$-th row of $W$ and $b_i$ be the $i$-th element of $b$, then we have

$$z_i = w_i^T x + b_i \quad (7)$$

During an adversarial attack, a noisy vector $\epsilon$ is generated and added to the input $x$, and then the output becomes:

$$z_{y,\epsilon} = w_{y,\epsilon}^T (x + \epsilon) + b_{y,\epsilon} \quad (8)$$

If the noise $\epsilon$ is small enough, assume that: $w_y \approx w_{y,\epsilon}$ and $b_y \approx b_{y,\epsilon}$. This assumption is valid if the "on/off" states of ReLU units do not change significantly when adversarial noise is added. Therefore

$$z_{y,\epsilon} \approx w_y^T x + b_y + w_y^T \epsilon = z_y + w_y^T \epsilon \quad (9)$$

Then, define NSR and apply Hölder's inequality:

$$NSR_y = \frac{|w_y^T \epsilon|}{|z_y|} \leq \frac{\|w_y\|_q \cdot \|\epsilon\|_p}{|z_i|} \quad (10)$$

where $\frac{1}{p} + \frac{1}{q} = 1$. In this work, we focus on L-infinity norm $\|\epsilon\|_\infty = \epsilon_{max}$, and therefore

$$NSR_y \leq \frac{\|w_y\|_1 \cdot \epsilon_{max}}{|z_y|} = R_2 \quad (11)$$

By combining the regularization term $R_2$ with Margin loss and Mean Square Error (MSE) loss for classification, the loss function for NSR Regularization is obtained, given by

$$loss = (z_y - 1)^2 + \sum_{i \neq y}(z_i - 0)^2$$
$$+ \sum_i max(0, 1 - z_y + z_i) + \beta log(R_2 + 1) \quad (12)$$

In the experiment, $\epsilon_{max}$ is set to 1, and $\beta$ is determined on the validation set. The Margin loss and the regularization term $log(R_2 + 1)$ are only used for correctly-classified samples; and for wrongly-classified samples, $loss$ only contains MSE loss. To help the model converge during the training process, the regularization term and Margin loss will not be added to the loss function until the 11-th epoch.

III. EXPERIMENT SETUP

*A. Data*

The ECG data in this experiment is from the China Physiological Signal Challenge 2018 (CPSC2018) [16], which is described in detail by [22]. This dataset is publicly available. There are 6877, 12-lead ECG recordings in this dataset. The ECG signals were sampled by a frequency of 500 Hertz, for 6 seconds to 60 seconds (note: a few signals have 144 seconds). Namely, each ECG sample (referring to the whole digit ECG signal from a patient) in the dataset has 12 leads/channels, and each sample has a variable length from 3000 to 72000. This dataset has 9 categories, which are "Normal" (918 samples), "Atrial fibrillation (AF)" (1098 samples), "First-degree atrioventricular block (I-AVB)" (704 samples), "Left bundle branch block (LBBB)" (207 samples), "Right bundle branch block (RBBB)" (1695 samples), "Premature atrial contraction (PAC)" (556 samples), "Premature ventricular contraction (PVC)" (672 samples), "ST-segment depression (STD)" (825 samples) and "ST-segment elevated (STE)" (202 samples).

We preprocessed the data. First, we removed the samples with multiple labels (477 samples). Second, from each class, we randomly put 5 samples into the validation set and 50 samples into the test set. In this way, we split the dataset into a training set of 5905 samples, a validation set of size 45, and a test set of size 450. Because this dataset is unbalanced, upsampling was done to augment the training set to be balanced in the number of samples in different categories. Third, for each sample, we removed Lead 3, 4, 5 and 6, which is due to the fact that Lead 3, 4, 5 and 6 are not independent of the rest of the leads and can be derived from Lead 1 and 2 [23]. Fourth, for each sample, we scaled each lead by its maximum absolute value to make each lead to be within the range of -1 to 1 [24]. Then, we randomly padded each sample with 0 on both ends. For a few samples with size larger than 33792, we discarded the signal elements after 33792. As a result, each sample after preprocessing is a tensor of shape $8 \times 33792$.

*B. Tuning NSR Regularization on the Validation Set*

To use the NSR Regularization, a proper β is needed to achieve the balance between robustness and accuracy, based on the performance on validation set. In Fig. 2, β in "βNSR" refers to the coefficient β of regularization term. From the result shown in Fig. 2, it can be seen that a too small β made the model not robust enough (e.g., "0.4NSR"), while a too large β led to a significant reduction in the F1 score on clean data (e.g., "1.1NSR" and "1.2NSR"). In this experiment, we chose "1.0" as the best value of β, because among these trials shown in Fig. 2, "1.0" is the largest value before significant reduction occurred in the F1 score on clean data.

## C. Tuning Jacobian Regularization on the Validation Set

To use Jacobian Regularization, a proper λ is needed to reach the balance between robustness and accuracy. We tried a series of λ on the validation set and selected a proper λ based on the performance. In Fig. 3, "λJacob" denotes Jacobian Regularization with coefficient λ. It can be seen that a too small λ made the model not robust enough (e.g., "4.0Jacob"), while a too large λ led to a significant reduction in the F1 score on clean data (e.g., "74.0Jacob" and "84.0Jacob"). In this experiment, we chose "44.0" as the best value of λ, because among these trials shown in Fig. 3, "44.0" is the largest value before the F1 score on clean data begins declining on validation set. The later experiment (in Fig. 4) also shows that clean accuracy of "54.0Jacob" indeed drops below 80% on test set, which is not acceptable in this experiment.

## IV. RESULTS AND ANALYSIS

In the experiment, we applied three defense methods to improve robustness of our customized CNN for classification of ECG signals from CPSC2018. We use PGD-100 attack and white noises to evaluate the performance of different methods. PGD-100 attack-based evaluation shows how a method performs against strong adversarial attack that could be lunched by an attacker perusing personal gain at the expanse of public health. White noise is much weaker than the adversarial noise from 100-PGD but could often exist in the real world. Fig. 4 and Fig. 5 show the results. In total, the CNN models trained by four methods are compared, which are: model trained with cross-entropy loss and clean data, denoted by "CE" (defenseless); model trained by Jacobian Regularization, denoted by "λJacob"; model trained by NSR Regularization, denoted as "βNSR"; model trained by vanilla adversarial training with 20-PGD, denoted by "advls_ϵ" where ϵ is the maximum noise level for training. The models were trained for 70 epochs. Training batch size is always 64. The noise levels in this experiment are always measured by L-infinity norm which can control the maximum noise amplitude. This experiment was conducted on a server with Nvidia Tesla V100 GPU processor (32 GB memory) and Intel(R) Xeon(R) E5-2698 v4 CPU processor (2.20GHz). Training "CE" costs less than 2 hours in total. Training with NSR Regularization costs about 600 seconds per epoch. Training with Jacobian regularization costs about 1200 seconds per epoch. Training with vanilla adversarial training costs about 600 seconds per epoch.

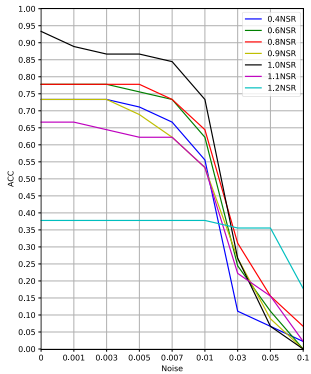
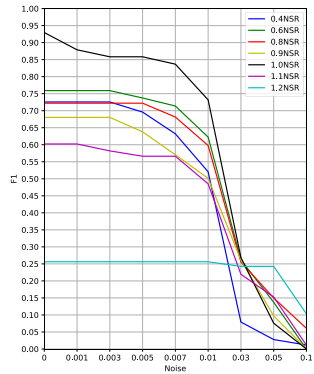

Fig. 2(a). Accuracy of NSR Regularization on the validation set

Fig. 2(b). F1 score of NSR Regularization on the validation set

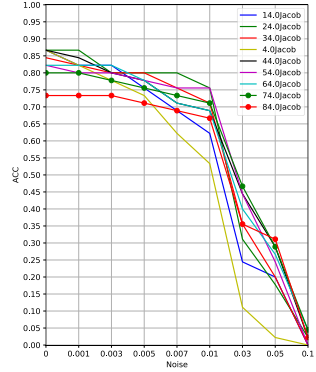
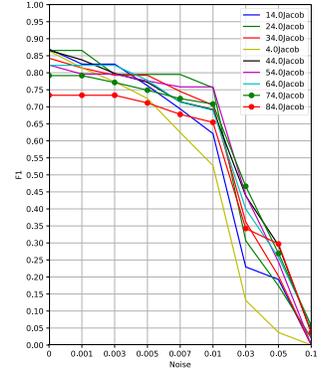

Fig. 3(a). Accuracy of Jacobian Regularization on the validation set

Fig. 3(b). F1 score of Jacobian Regularization on the validation set

## A. 100-PGD Evaluation

In this section, we evaluate the trained models by using 100-PGD adversarial attack. We focus on noise level no larger than 0.01. When noise level is larger than 0.01, the perturbation is so significant that even human eyes can hardly recognize what the original image should be and thus it is no longer a noise-robustness problem (see Appendix A).

First, the "CE" version of our customized CNN can classify this ECG data with average accuracy of 80% and F1 score of nearly 0.790 on clean data. Compared with results of other entries reported in CPSC2018 [16], this F1 score is in top 6 (F1 scores of the Top-6 methods are 0.837, 0.830, 0.806, 0.802, 0.791 and 0.783), which means the "CE" version reached a satisfying accuracy for this classification task on clean data, without using recurrent neural network. Second, according to Fig. 4, compared with "CE", three defense methods successfully improve the robustness of the CNN against 100-PGD adversarial attack. 100-PGD attack is so strong that the accuracy of "CE" drops to 0% at noise level of 0.01. As a comparison, at the noise level of 0.01, "1.0NSR", "44.0Jacob" and "advls_0.01" keep an average accuracy of more than 60% and F1 score of more than 0.6, ranking at least top 30 in CPSC2018. Furthermore, from Fig. 4, it is clear that "1.0NSR" and "advls_0.01" have higher average accuracy about 83% and F1 score about 0.83 on clean data, which would rank in top-3 in CPSC2018. The increase in clean accuracy is achieved by using the regularization (NSR Regularization) or data augmentation (adversarial samples generated by vanilla adversarial training). This improvement is significant in this challenging experiment due to the small sample size and high-dimensional data samples. The results also suggest that improving robustness may not always be in conflict of increasing accuracy. We note that none of the methods in CPSC2018 are evaluated against 100-PGD adversarial attack.

However, we observe that "advls_ϵ" has some weaknesses. First, it is very sensitive to the user-defined noise level ϵ, which makes the CNN to be robust only around the specific noise level of ϵ. From Fig. 4, "advls_0.01" has a good resist to 100-PGD at noise level of no larger than 0.01 and the accuracy drops significantly after that; For "advls_0.05", accuracy drops slowly before 0.05 and then significantly. Also, a too large ϵ leads to significant reduction of classification accuracy on clean data (e.g., "advls_0.05" and "advls_0.1"). This can be explained as follows. A too large perturbation on the sample can push the input to cross the optimal decision boundary for classification. A too small $\epsilon$ all has relatively week effect on CNN robustness (e.g., "advls_0.005").

In this evaluation, we focus on noise level no larger than 0.01. Within this noise level, "1.0NSR" outperforms "44.0Jacob" on clean data accuracy, and it also outperforms "advls_0.01" on robustness against 100-PGD attack.

### B. White Noise Evaluation

In this section, we evaluate the defense methods using white noises of different levels. Basically, white noises from a uniform distribution are added to the samples in the test sets, and the accuracy of each trained model on these noisy samples are measured. The L-infinity norm of the random noise can be controlled in each individual trail. Compared with 100-PGD, random white noise is more like the natural noises in ECG signals, which are originated from the recording device (e.g. electromyogram noise and power line interference). The results are shown in Fig. 5. Obviously, two regularization-based methods, "1.0NSR" and "44.0Jacob", outperform vanilla adversarial training by making the CNN robust to white noises. According to Fig. 5, "1.0NSR" has the best performance when noise level is smaller than 0.4 and keeps the accuracy almost unchanged when noise level is smaller than 0.2. When the noise level exceeds 0.4, the accuracy starts to drop significantly, which may be related to the strong assumption in the NSR Regularization method. However, 0.4 is a very large noise level, under which the signals are changed significantly (see noise level of 0.1 at Appendix A). As a result, "1.0NSR" has enough robustness against white noises. "44.0Jacob" in general has a very good resistance to large white noises and an average accuracy of more than 70% at noise level of 0.6. However, the accuracy of every "advls_$\epsilon$" drops significantly as the white noise level increases. The performance of ""advls_$\epsilon$" in this evaluation is not as good as that in the 100-PGD evaluation. The reason could be that the vanilla adversarial training is based on the specific type of adversarial noises from PGD, while regularization term-based methods are not.

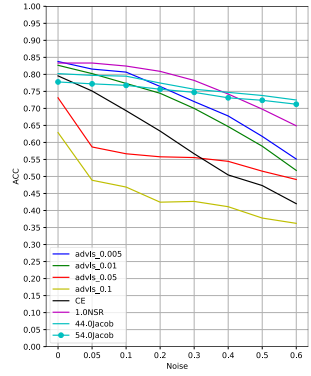
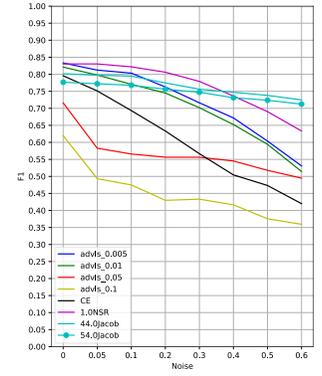

Fig. 5(a). Accuracy of methods on the test set, under white noise

Fig. 5(b). F1 score of methods on the test set, under white noise

### V. CONCLUSION

In this paper, for classification of the ECG signal data from the China Physiological Signal Challenge 2018, we designed a CNN that can handle high-dimensional and variable length input. We applied three representative defense methods to improve the robustness of this CNN against noises. 100-PGD adversarial attack and white noises are used to evaluate the defense methods, and all of the methods successfully improved the robustness of CNN against 100-PGD and white noise, compared to the model trained only with cross-entropy loss and clean data. We found out that vanilla adversarial training is very sensitive to the user-defined noise level $\epsilon$, and the regularization-based defense methods can provide better resistance against white noises than vanilla adversarial training. We hope that this study could facilitate the development of robust solutions for automated ECG diagnosis.

Note:
(1) The implementation of this experiment is available at https://github.com/SarielMa/ICMLA2020_12-lead-ECG
(2) All figures are in high resolution, please zoom in.

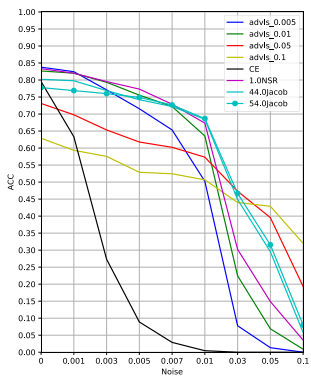
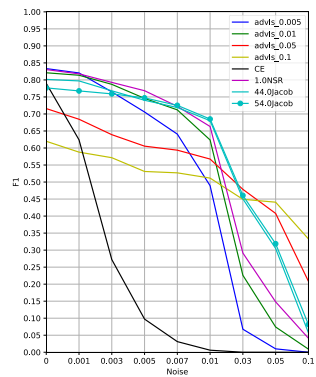

Fig. 4(a). Accuracy of methods on the test set, under 100-PGD

Fig. 4(b). F1 score of methods on the test set, under 100-PGD

APPENDIX A

In this section, Lead 1 of an ECG signal with different levels of adversarial noises is shown. The adversarial attack is PGD-100 and the target network is our CNN, which has been trained with cross entropy Loss for 70 epochs. As we can see, when the noise level is larger than 0.01 (Fig. 6 (g)), the original ECG signal is even hardly recognizable by human eye.

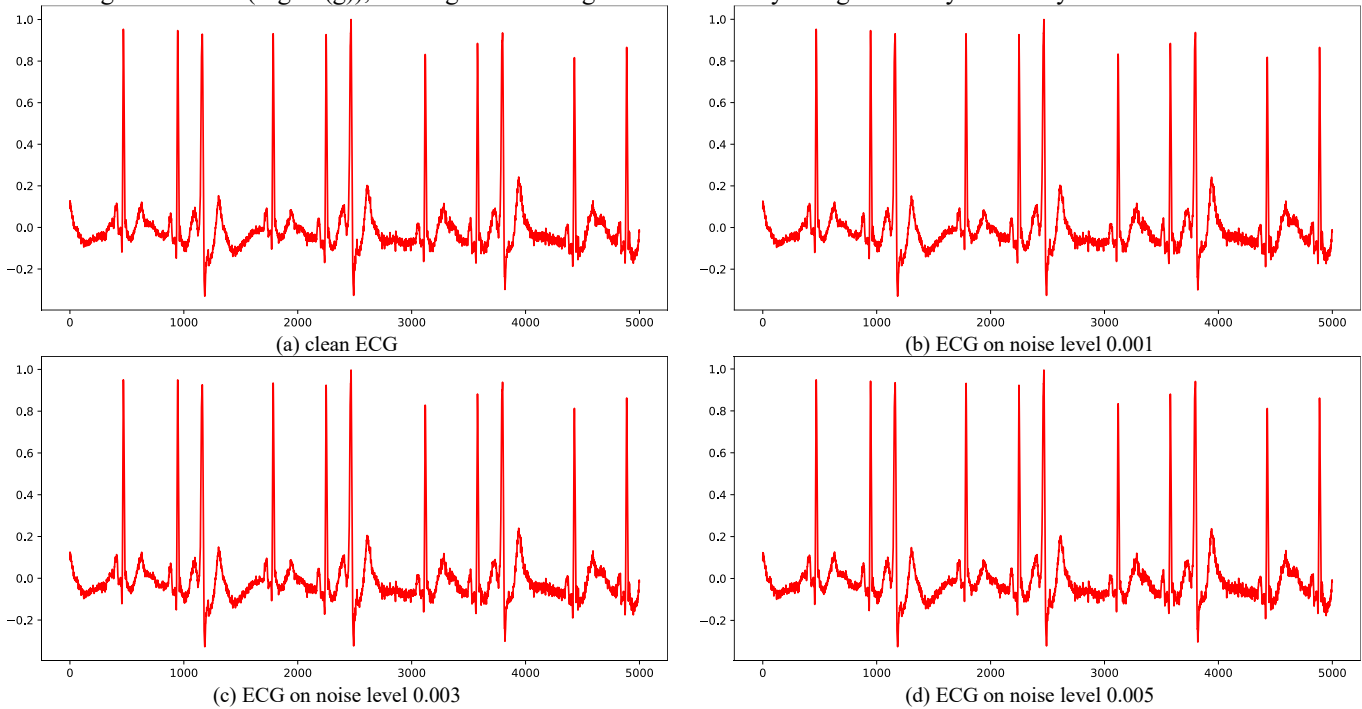

(a) clean ECG

(b) ECG on noise level 0.001

(c) ECG on noise level 0.003

(d) ECG on noise level 0.005

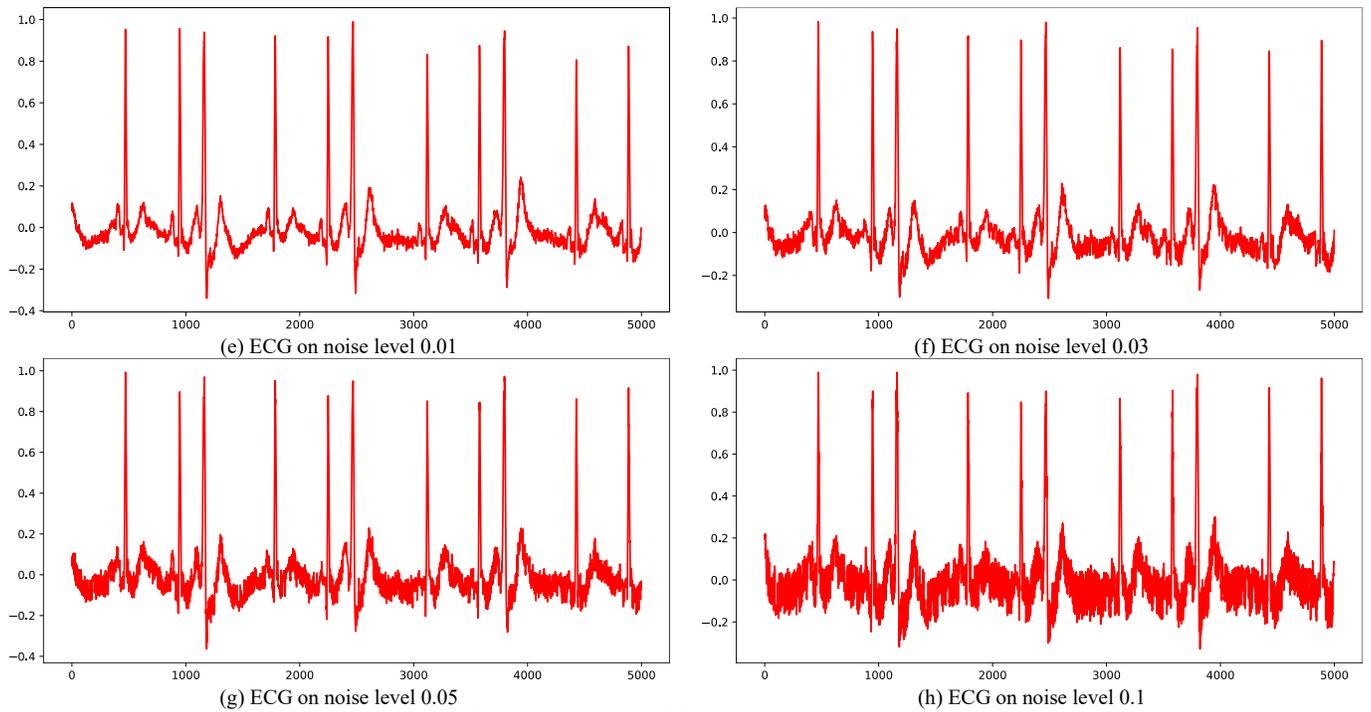
Fig. 6. ECG signal Lead I on different levels of adversarial noises

## APPENDIX B

The tables in the section are results shown in Fig. 4.

TABLE 1. ACCURACY OF METHODS ON THE TEST SET, UNDER 100-PGD

| noise level | 0 | 0.001 | 0.003 | 0.005 | 0.007 | 0.01 | 0.03 | 0.05 | 0.1 |
|---|---|---|---|---|---|---|---|---|---|
| advls_0.005(ACC) | 0.840 | 0.820 | 0.770 | 0.720 | 0.650 | 0.500 | 0.080 | 0.010 | 0.000 |
| advls_0.01(ACC) | 0.830 | 0.820 | 0.790 | 0.760 | 0.720 | 0.640 | 0.220 | 0.070 | 0.010 |
| advls_0.05(ACC) | 0.730 | 0.700 | 0.650 | 0.620 | 0.600 | 0.570 | 0.470 | 0.400 | 0.190 |
| advls_0.1(ACC) | 0.630 | 0.590 | 0.580 | 0.530 | 0.520 | 0.510 | 0.440 | 0.430 | 0.320 |
| CE(ACC) | 0.800 | 0.630 | 0.270 | 0.090 | 0.030 | 0.000 | 0.000 | 0.000 | 0.000 |
| 1.0NSR(ACC) | 0.830 | 0.820 | 0.800 | 0.770 | 0.730 | 0.670 | 0.300 | 0.150 | 0.040 |
| 44.0Jacob(ACC) | 0.800 | 0.800 | 0.770 | 0.740 | 0.720 | 0.680 | 0.450 | 0.300 | 0.060 |
| 54.0Jacob(ACC) | 0.780 | 0.770 | 0.760 | 0.750 | 0.730 | 0.690 | 0.470 | 0.320 | 0.080 |

TABLE 2. F1 SCORES OF METHODS ON THE TEST SET, UNDER 100-PGD

| noise level | 0 | 0.001 | 0.003 | 0.005 | 0.007 | 0.01 | 0.03 | 0.05 | 0.1 |
|---|---|---|---|---|---|---|---|---|---|
| advls_0.005(F1) | 0.833 | 0.820 | 0.765 | 0.706 | 0.641 | 0.489 | 0.068 | 0.010 | 0.000 |
| advls_0.01(F1) | 0.821 | 0.814 | 0.787 | 0.746 | 0.712 | 0.624 | 0.226 | 0.074 | 0.009 |
| advls_0.05(F1) | 0.716 | 0.684 | 0.639 | 0.605 | 0.594 | 0.568 | 0.479 | 0.408 | 0.209 |
| advls_0.1(F1) | 0.620 | 0.588 | 0.572 | 0.531 | 0.527 | 0.512 | 0.449 | 0.441 | 0.333 |
| CE(F1) | 0.790 | 0.624 | 0.272 | 0.097 | 0.031 | 0.006 | 0.000 | 0.000 | 0.000 |
| 1.0NSR(F1) | 0.830 | 0.818 | 0.793 | 0.768 | 0.722 | 0.663 | 0.292 | 0.148 | 0.041 |
| 44.0Jacob(F1) | 0.802 | 0.797 | 0.768 | 0.740 | 0.719 | 0.680 | 0.451 | 0.305 | 0.057 |
| 54.0Jacob(F1) | 0.777 | 0.768 | 0.759 | 0.747 | 0.725 | 0.685 | 0.461 | 0.319 | 0.082 |